\documentstyle[12pt,a4]{article}

\begin{document} 
\bibliographystyle{unsrt} 
\baselineskip= 16pt 

\begin{titlepage}
\title{Low-Temperature Features of Nano-Particle Dynamics}
\author{R. Sappey, E. Vincent, M. Ocio and J. Hammann\\ Service de Physique de l'Etat 
Condens\'e, CEA Saclay,\\ 91191 Gif sur Yvette Cedex, France \\ \\
}
\maketitle
\begin{abstract}
In view of better characterizing possible quantum effects in the dynamics of nanometric particles, we measure the effect on the relaxation of a slight heating cycle. The effect of the field  amplitude is studied; its magnitude is chosen in order to induce the relaxation of large particles ($\sim 7nm$), even at very low temperatures ($100mK$). Below  $1K$, the results significantly depart from a simple thermal dynamics scenario.
\end{abstract}


\vskip 1.0 cm
\noindent
E-mail: sappey@spec.saclay.cea.fr, vincent@spec.saclay.cea.fr
\vskip 0.5 cm
\noindent
{\em Submitted for the Proceedings of ICM97}

\end{titlepage}

The contribution of quantum tunneling events \cite{chud} to the reversal of the magnetic moment of nano-particles is commonly investigated by relaxation measurements \cite{tejada}. However, the measured relaxation rate $S$  (``magnetic viscosity'', defined as the time-logarithmic derivative of the magnetization) is an intricate combination of the effective temperature $T^*(T)$  
 (involved in the activated crossing of a barrier U)  with the distribution of anisotropy barriers $P(U)$.
$T^*(T)$ is the quantity of interest; it is equal to $T$ for thermal dynamics, and should level off to some low-T plateau for quantum dynamics \cite{chud} (sketch in the inset of Fig.2). The knowledge of $P(U)$ is highly uncertain  for the smallest barriers which are explored at the lowest temperatures, where precisely one aims at extracting $T^*(T)$.

Thermal dynamics is strongly accelerated by a temperature increase, wehereas quantum processes should be much less affected; in this spirit, we have studied \cite{eplnous}  the fraction of the relaxation rate at a given $T$ which survives to a stage at a slightly higher temperature $xT$.
After the field change at $T$ which triggers the relaxation ($t=0$), we heat the sample to $xT$ during 200s ($1\le x \le 3$), and then measure the viscosity at $T$ (see Fig.1). We  normalize this viscosity  to that measured without heating to $xT$. 
In the obtained ``Residual Memory Ratio" (RMR, \cite{eplnous}), $P(U)$  is (to  first order) eliminated. 
More precisely, we have computed this ratio for some example distributions $P(U)$ (which we consider temperature independent, as expected for independently relaxing particles). Fig.2  presents the results; remarkably, $RMR(x)$ is nearly insensitive to extreme choices of $P(U)$, while clearly picturing the thermal or quantum nature of the dynamics. 

We have previously applied a similar measurement technique \cite{eplnous} to a sample of $\gamma -Fe_2O_3$ particles diluted in a silica matrix with a  $4.3\,10^{-4}$ volume fraction \cite{sample}.  The particle sizes correspond to a log-normal distribution of $d_0\sim 7nm$ and $\sigma \sim 0.3$. The relaxations were 
induced by switching off a 62 Oe field in which the sample had been cooled from 6K. For $0.06K\le T\le 1K$, the measured $RMR(x)$ clearly departs from the thermal expectation (see details in \cite{eplnous}).

The main drawback of such measurements at fixed low field is that
the relaxing objects at very low T are
 by far smaller than the typical size of $7nm$, and cannot be clearly identified; one may  wonder about the contribution of surface excitations in larger particles \cite{Berkowitz}.  In order to measure  quasi-macroscopic particles (close to the peak of the size distribution), we have performed measurements using higher fields. Due to the important flux drifts caused by this procedure, we  study a less diluted sample of the same particles as before, with a $3.4\,10^{-3}$ volume fraction; the dipolar fields ($H_{dip}=4\pi M/V\sim {\rm a\  few\ } Oe$) remain small compared to the anisotropy field $H_a\sim 1000Oe$).

We prepare the sample at constant temperature in a negative field {$H_0=-2000Oe$}, in which almost all anisotropy barriers are overcome; then we sweep this field to some value  $H_1$ ($-100Oe<H_1<1500 Oe$). Small particles relax, but the larger ones remain negatively oriented; we wait 1 hour, mainly for damping the drift of the magnet.
  Then we increase the field from $H_1$ to $H_2$ (here we keep $H_2-H_1$ equal to $46Oe$). This last field increase triggers the relaxation of a set of particles of the corresponding size; at several given temperatures T, we have studied the relaxation rate $S(T,H_2)$ as a function of $H_2$. The results are shown in Fig.3, where at each temperature a peak in the relaxation rate can be seen, reflecting the distribution of the field-modulated barriers (see a discussion of such $S(T,H)$ in \cite{sampaio}). At 100mK,
we observe during our procedure a transitory heating which is related to the large change in the sample magnetization.
 This difficulty has already been noted in the very few existing experiments at high field and very low T \cite{paulsen}, and related to avalanche processes. 
At 100 mK, the effect is important for $H_2>500Oe$; the right-hand part of this curve is therefore artificially depressed (Fig.3). At 300mK the effect is already much weaker \cite{romain}.

Such a stray heating impedes a simple interpretation of viscosity measurements,
the distribution of objects which are about to relax being modified by the initial temperature increase. However,  
 if this initial state is reasonably reproducible, it will not affect an $RMR$-type measurement, in which the initial distribution is of very little influence (Fig.2). 
We have checked this point; our viscosity values at 100mK remain reproducible within $\pm 20\%$, and within $\sim \pm15-5\%$ at higher temperatures. 
Besides, at 100 and 300mK, we have measured $RMR(x)$ for two different values of $H_2$; 
different sizes of particles are relaxing in each case, but the $RMR(x)$ decrease with $x$ remains about the same. The $H_2$ choices at different temperatures (made for practical reasons \cite{romain}) are therefore of no significant influence on 
our $RMR(x)$ discussion.

The results are displayed in Fig.4. From 1K and below, $RMR(x)$ clearly departs from the simple thermal expectation, which is verified at 3K. As was the case in our previous low-field results \cite{eplnous}, no plateau of $RMR(x)$ is visible, suggesting a more complex behavior than pictured for the quantum case in Fig.2. If the $RMR$-anomalies are to be ascribed to quantum mechanics, then one must think of a distribution of thermal-to-quantum crossover temperatures among the particles. 
This may be due to a distribution of anisotropy energy densities; for nanometric  particles, the anisotropy arises from multiple origins (shape, surface...). 
On the other hand, if the behavior is purely thermal, then the observed anomalies imply that the very common description that we use here is not correct; the T-independence of $U$ and $P(U)$ may be questioned, which means for instance examining how far the particles are rigid macro-moments, or how independent of each other they can be considered, even in highly diluted samples.

\newpage

\newpage

\pagestyle{empty}
\noindent{\Large \bf Figure captions}
\vskip 1 truecm
\underline {Figure 1 :}  Temperature and magnetic moment variations, recorded during a  $RMR$ procedure at 3K (dashed line: expectation for the quantum case \cite{chud}).

\vskip 0.5 cm
\underline{Figure 2 :} Calculated $RMR(x)$, for very different $P(U)$  choices, in the thermal (i) and quantum (ii) cases.

\vskip 0.5 cm
\underline{Figure 3 :} Measured viscosity as a function of the final field value $H_2$ (see procedure in the text), at different temperatures.

\vskip 0.5 cm
\underline{Figure 4 :} Measured $RMR(x)$, at different temperatures (the negative values are compatible with zero within the error bars). The solid line is the expectation for a simple thermal dynamics scenario.

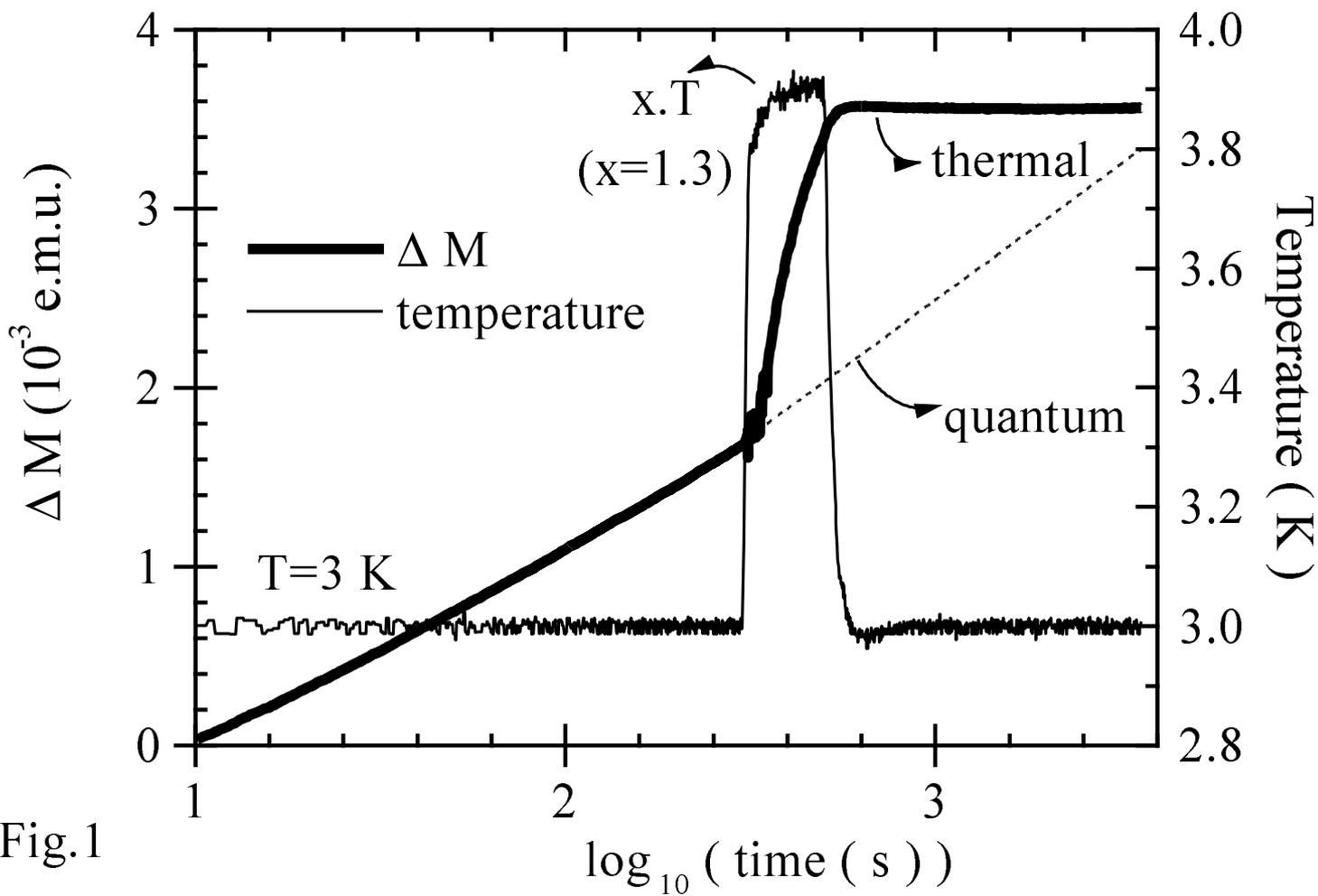
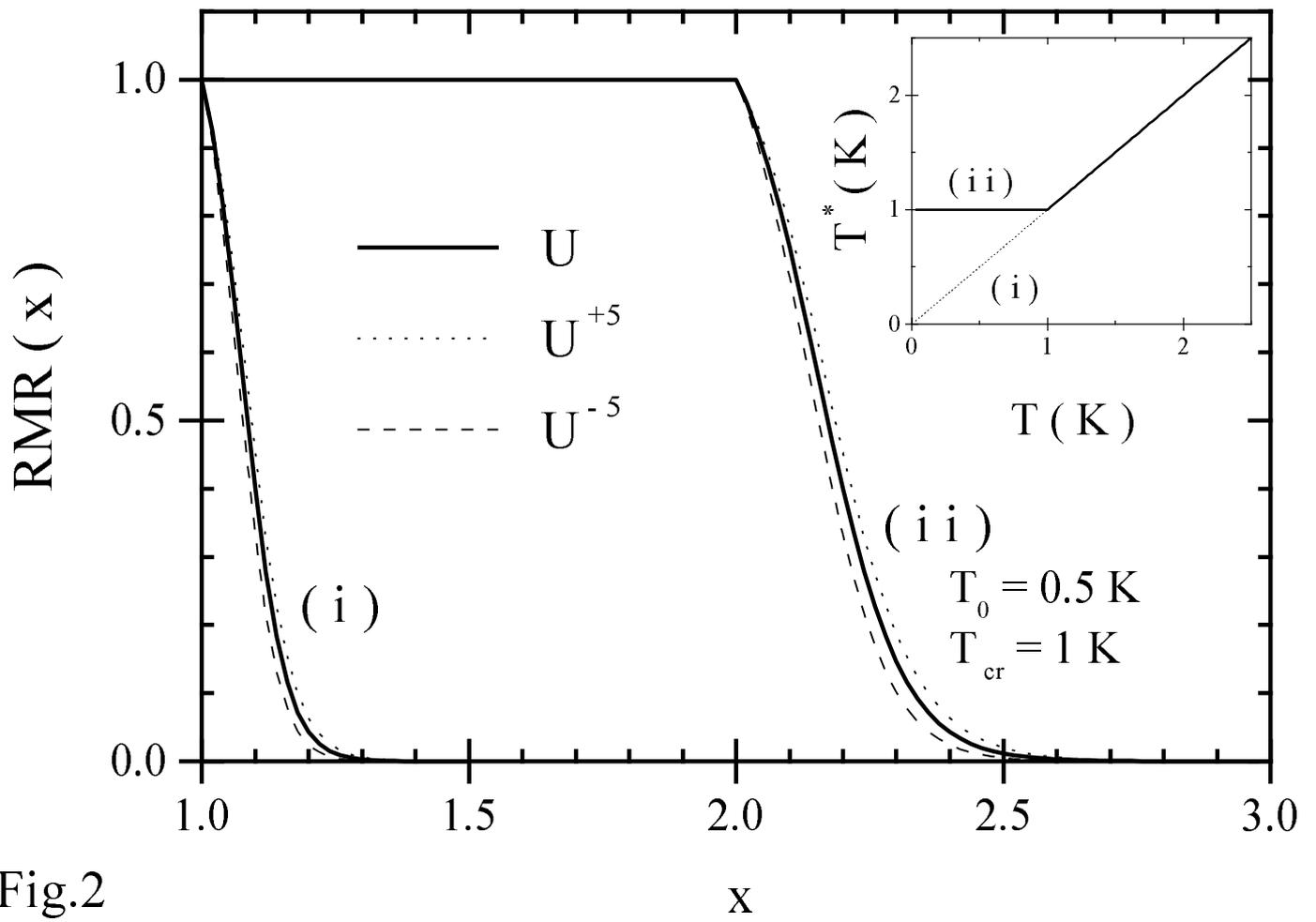
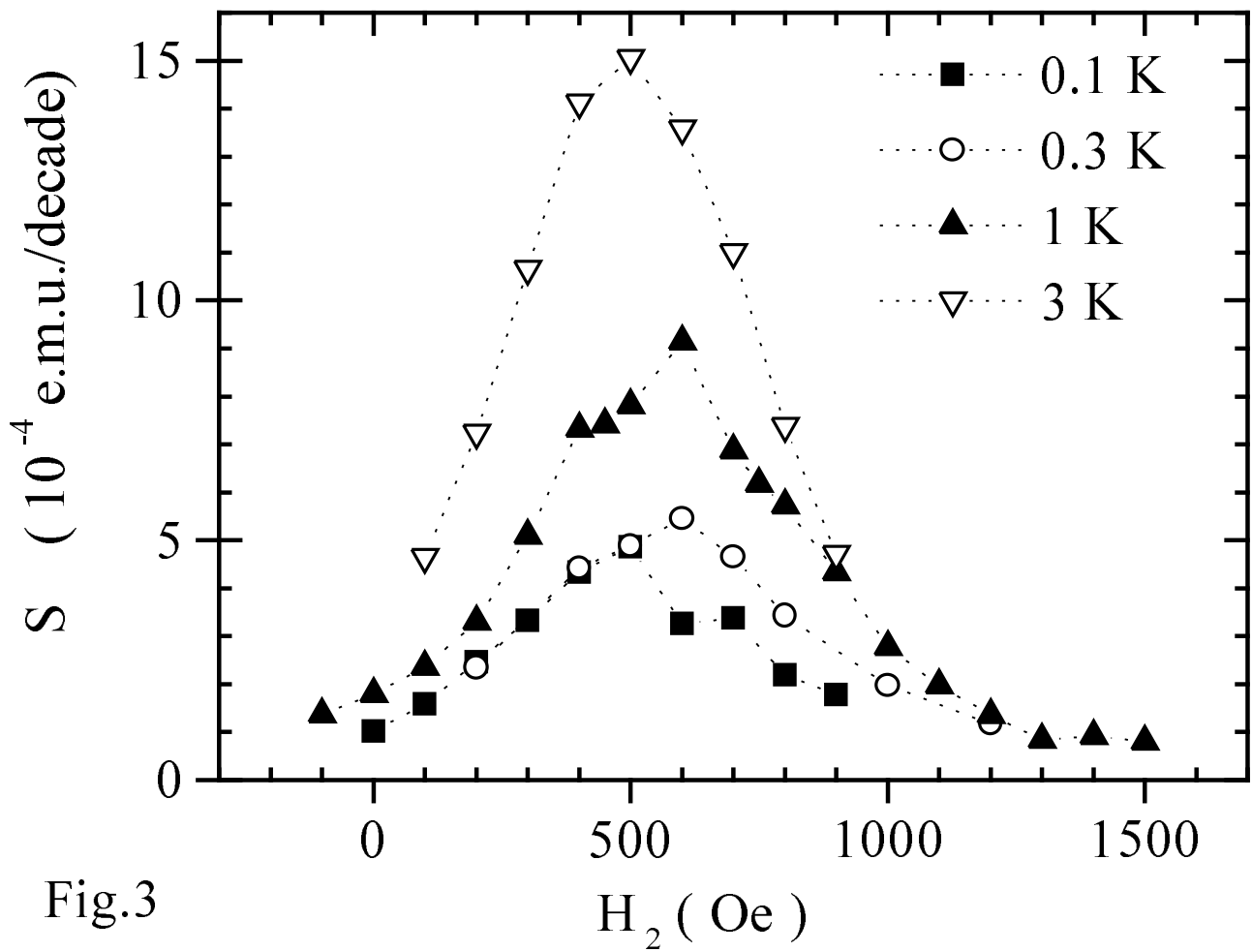
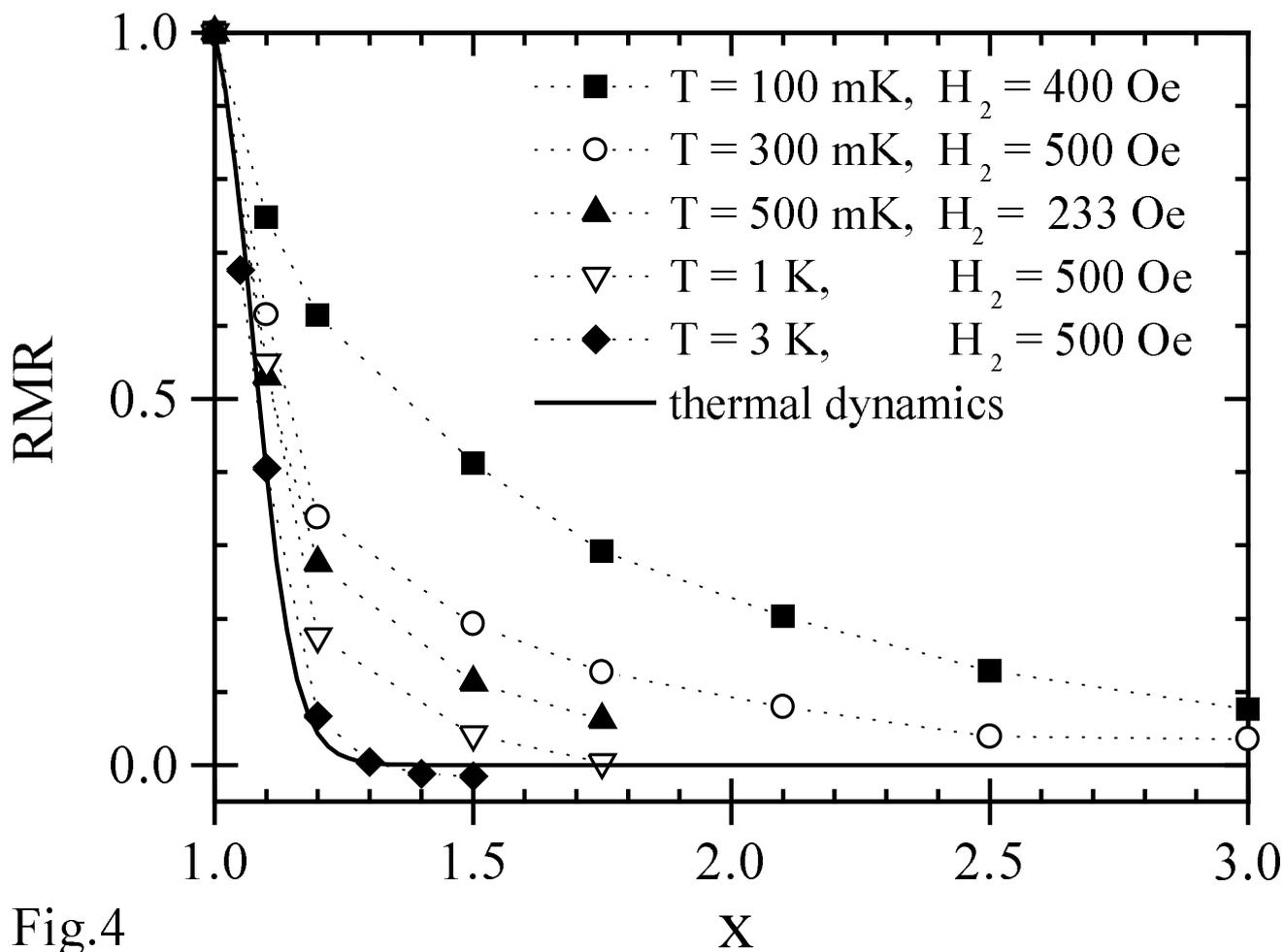
\end{document}